Amin Alibakhshi[1], Weiqiu Chen[2,3], Michel Destrade[3,4]

# Nonlinear vibration and stability of a dielectric elastomer balloon based on a strain-stiffening model

[1]Department of Mechanical Engineering, Science and Research Branch, Islamic Azad University, Tehran, Iran. Email: alibakhshiamin@yahoo.com
[2]Shenzhen Research Institute of Zhejiang University, Shenzhen 518057, P. R. China
[3]Key Laboratory of Soft Machines and Smart Devices of Zhejiang Province & Department of Engineering Mechanics & Soft Matter Research Center, Hangzhou 310027, P. R. China
[4]School of Mathematical and Statistical Sciences, NUI Galway, University Road, Galway, Republic of Ireland. Email: michel.destrade@nuigalway.ie

**Abstract**

Limiting chain extensibility is a characteristic that plays a vital role in the stretching of highly elastic materials. The Gent model has been widely used to capture this behaviour, as it performs very well in fitting stress-stretch data in simple tension, and involves two material parameters only. Recently, Anssari-Benam and Bucchi [Int. J. Non. Linear. Mech. 2021, 128, 103626] introduced a different form of generalised neo-Hookean model, focusing on the molecular structure of elastomers, and showed that their model encompasses all ranges of deformations, performing better than the Gent model in many respects, also with only two parameters. Here we investigate the nonlinear vibration and stability of a dielectric elastomer balloon modelled by that strain energy function. We derive the deformation field in spherical coordinates and the governing equations by the Euler-Lagrange method, assuming that the balloon retains its spherical symmetry as it inflates. We consider in turn that the balloon is under two types of voltages, a pure DC voltage and a DC voltage superimposed on an AC voltage. We analyse the dynamic response of the balloon and identify the influential parameters in the model. We find that the molecular structure of the material, as tracked by the number of segments in a single chain, can control the instability and the pull-in/snap-through critical voltage, as well as chaos and quasi-periodicity. The main result is that balloons made of materials exhibiting early strain-stiffening effects are more stable and less prone to generate chaotic nonlinear vibrations than softer materials, such as those modelled by the neo-Hookean strain-energy density function.

# 1 Introduction

Electroactive polymers are smart materials that experience deformation in response to external electric loads. The past two decades have seen a growing trend toward studying, characterizing, and modelling electroactive polymers and finding exciting new applications. Many attempts have been made to develop this kind of smart materials.

Electroactive polymers are classified into several groups such as ferroelectric polymers [1–7], electro-strictive graft polymers [8, 9], electro-rheological fluids [10–14], ionic polymer-metal composites [15–20], stimuli-responsive gels [21–25], and dielectric elastomers (DEs) [26, 27]. DEs are important for a wide range of scientific and industrial processes, as they can be used as smart structures that exhibit large deformations under electrical loadings. In real-world applications and academic studies, various types of DE structures have been introduced and designed, including DE balloons [28], DE tubes [29], DE beams [30], and DE plates [31, 32]. These prototypes can be used as actuators, sensors, and/or energy harvesters, and find diverse applications such as pumps, haptic devices, artificial muscles, and adaptive lenses.

Vibration analysis and stability analysis of DE balloons have been the object of intense research efforts for many years, because of their practical relevance for prototypes. Hence, Zhu et al [33] analysed the nonlinear vibration of a DE spherical shell using the neo-Hookean strain energy function, and obtained time history diagrams and frequency-amplitude diagrams. They also calculated the natural frequency of soft balloons around equilibrium stretches. Yong et al. [34] studied the nonlinear dynamic response of a thick-walled neo-Hookean DE balloon under static and sinusoidal voltages. Jin and Huang [35] assessed the random vibrations of a DE neo-Hookean balloon, obtained the random equation around equilibrium stretches, and solved the problem using the stochastic averaging technique. Alibakhshi and Heidari [36] used the multiple scales method to analyse the nonlinear resonance of a DE neo-Hookean spherical shell and derived the frequency-amplitude curve and time responses. Tang et al. [37] addressed analytically the nonlinear free oscillation of a neo-Hookean DE balloon with the Newton–harmonic balance method. Liu and Zhou [38] employed the shooting method and the arc-length continuation method to find periodic responses and explore the nonlinear resonance of a DE neo-Hookean balloon. Tang et al [39] developed a vibration and stability analysis for a spherical shell made of neo-Hookean DE, using the Newton–harmonic balance method.

These studies were all based on the neo-Hookean model for the balloon, a model that cannot capture the highly nonlinear strain-stiffening effect observed experimentally in the large deformations of DEs, and caused by limiting chain extensibility. As an alternative, the Gent strain energy function [40] is often employed to model DEs and take limiting chain extensibility into account. Hence Chen and Wang [41] used the Gent model when they explored the nonlinear dynamic characteristics of a DE balloon. Chen et al [42] studied the electromechanical instability of a Gent DE balloon. Lv et al [43] investigated nonlinear vibrations of a Gent DE balloon, taking damping effects into account. They conducted their dynamic analysis by depicting time-stretch responses, phase-plane diagrams, and Poincaré maps. Deng and Li [44] analysed the influence of a protective passive layer on the vibrational behaviour of a DE sphere, also using the Gent model. Mao et al. [45] investigated small-amplitude free vibrations of pressurized electro-active Gent spherical balloons using incremental equations of motion, and found that both the internal pressure and the radially applied voltage difference can be used effectively to tune the balloon's vibration behavior. Liang and Cai [46] identified new electromechanical instabilities in Gent DE balloons, namely a localized bulging-out.

Another model based on the molecular structure of rubbers and polymers is the Arruda-Boyce model [47], which also captures the limiting chain extensibility effect. The Arruda-Boyce model is based on statistical mechanics, and reduces to the Gaussian network-based neo-Hookean model in the small-deformation regime. The literature using this model to investigate the response of DEs is not extensive on account of the complexity in its mathematical formulation, which involves the inverse Langevin function (sometimes a truncated Taylor series expansion of that function is adopted, but then the resulting polynomial cannot capture the limiting extensibility effect). An example is the work of Itskov et al.[48], who studied the voltage-stretch response of a DE experimentally and theoretically using the eight-chain

Arruda–Boyce model. The Gent model is an accurate and simple approximation to the Arruda-Boyce model which accounts for limiting chain extensibility and lends itself to analytical results.

However, the Gent model cannot predict an accurate response and capture all ranges of deformations at the same time. To overcome this limitation, extended versions of the Gent model have been proposed over the years. Among them, the Gent-Gent model of Pucci and Saccomandi [49] is the most versatile model, and Mangan and Destrade [50] used it to model the inflation of elastic tubes and balloons. Alibakhshi and Heidari [51] investigated the nonlinear dynamics of a soft Gent-Gent DE spherical shell; they concluded that the inclusion of the Gent-Gent model can suppress chaos in DE balloons.

Recently, Anssari-Benam and Bucchi [52] developed a simple hyperelastic model, also a generalised neo-Hookean model in the sense that its strain energy density depends only on the first principal invariant of strain, like the Gent model. They reported that, like the Gent model, their model can incorporate limiting chain extensibility characteristics. It displays further advantages such as being able to describe many ranges of deformation and having a simple mathematical form, with only two material constants, naturally related to a molecular chain description of soft and elastic materials, providing a sound structural basis [50] and improved fitting compared to the Gent model [52]. Their model belongs to a subclass of the more general model proposed by Davidson and Goulbourne [53], developed to capture the behaviour of soft polymeric films with chain entanglements and crosslinks.

With this paper, we argue that understanding the performance of DEs balloons based on this type of constitutive model is important, especially in the regime of nonlinear dynamics and stability, where molecular chains are extended and contribute to a sharp stiffening of the membrane. We shall thus extend the results of Anssari-Benam et al. [52], who recently used that model for elastic instabilities of balloons (and tubes) inflated quasi-statically, to now include electro-mechanical coupling, dynamic loading and nonlinear vibrations. We also extend the recent results of Khurana et al.[54], who recently used the Davidson and Goulbourne [53] model to study the nonlinear vibrations of DE flat plates.

In Section 2 we derive the governing equation of motion using the Euler-Lagrange equation for a purely radial large deformation of the spherical shell, considering that the spherical symmetry holds. In Section 3, we study the static and dynamic stability of the solution of the system. We find that early strain-stiffening effect (corresponding to a small number of chain elements) stabilises the balloon. In Section 4, we report numerical results and discuss the resulting time traces plots, phase-plane diagrams, Poincaré sections, and Fast Fourier transforms (FFTs). In Section 5 we summarise the main conclusions and present avenues for possible future work.

## 2 Mechanical modelling

The schematic representation of a DE balloon is depicted in Fig. 1. The balloon undergoes large deformation and is modelled by finite strain theory. In the reference configuration, before external loadings are applied, the thickness and radius of the balloon are denoted by $H$ and $R$, respectively. Once the balloon is subjected to a tensile load $P$ and a voltage $V$, the current configuration is used where the thickness and radius become $h$ and $r$. The thicknesses are small enough to assume that the membrane hypothesis holds (See Mangan and Destrade [50] for details on the validity of this assumption).

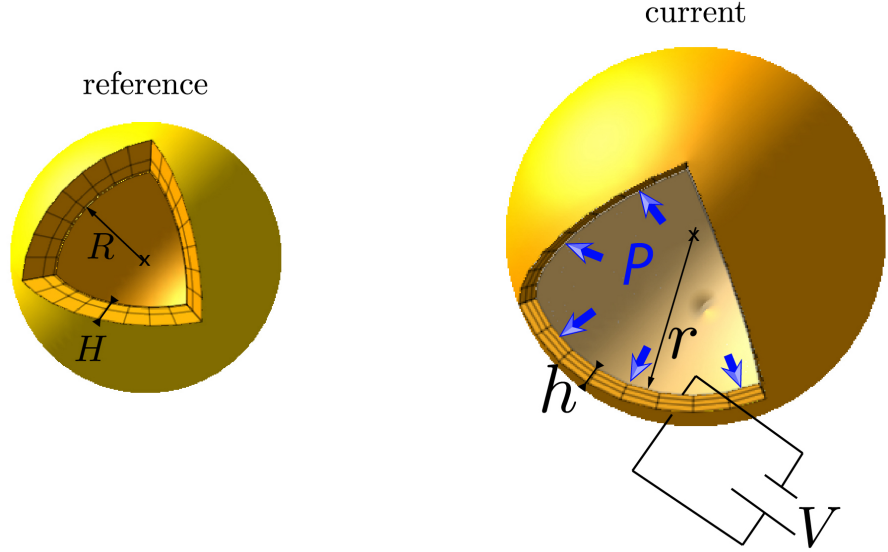

**Figure 1. Schematic view of the dielectric elastomer balloon in the reference and current configurations. Note that the thicknesses are amplified in the figure for illustration purposes. In the analysis, they are assumed to be infinitesimal compared to the radii, leading to the membrane approximation** [50]**.**

In line with previous studies, the vibration of a spherical balloon made of DE is described in terms of the principal stretches of the elastomer. We assume that that the material is isotropic and incompressible, that the membrane undergoes a purely radial deformation, and that the balloon retains its spherical symmetry as it inflates. We do not consider the possibility of non-spherical bulging, an inhomogeneous deformation which is beyond the remit of this study.

The radial and circumferential stretches for the thin spherical shell are expressed as

$$\lambda_r = \frac{dr(t)}{dR}, \qquad (1a)$$

$$\lambda_\theta = \lambda_\phi = \frac{r(t)}{R}, \qquad (1b)$$

respectively, and incompressibility imposes that

$$\frac{dr(t)}{dR} = \frac{R^2}{r^2}. \qquad (2)$$

Calling the circumferential stretch $\lambda = \lambda_\theta = \lambda_\phi$ (and then the radial stretch is $\lambda_r = \lambda^{-2}$), we now derive the governing equations of motion in terms of $\lambda$ and its derivatives.

Because $r = \lambda(t)R$, the kinetic energy of the thin spherical shell is expressed as

$$U_k = 2\pi R^4 H \rho \left(\frac{d\lambda}{dt}\right)^2, \qquad (3)$$

where $\rho$ is the mass density of the material (it is constant because of incompressibility).

The work done by the inflation pressure is

$$W_P = \frac{4}{3}\pi PR^3(\lambda^3 - 1), \tag{4}$$

so that the associated mechanical potential energy is

$$U_P = -\frac{4}{3}\pi PR^3\lambda^3 + const. \tag{5}$$

The strain energy function proposed by Anssari-Benam and Bucchi [52] on the basis of statistical mechanics of molecular chains is

$$W^{ABB} = \mu N\left[\frac{1}{6N}(I_1 - 3) - \ln\left(\frac{I_1 - 3N}{3 - 3N}\right)\right], \tag{6}$$

where $N$ is the number of straight segments in a single chain of elastomers (an integer, also known as the number of Kuhn segments), and $\mu$ is a constant, related to $\mu_0$, the infinitesimal shear modulus, by the relation $\mu = 3\mu_0 \frac{N-1}{3N-1}$. Note that the constraint $I_1 < 3N$ must apply for the natural logarithm to be defined. Note also that for large $N$, the material behaves as a neo-Hookean solid, with strain energy $W^{nH} = \mu_0(I_1 - 3)/2$.

In Eq. (6), $I_1$ denotes the first invariant of the right Cauchy-Green deformation tensor; here, it is equal to $I_1 = \lambda_r^2 + \lambda_\theta^2 + \lambda_\phi^2 = 2\lambda^2 + \lambda^{-4}$. It follows that the stretch cannot be greater than $\lambda_m$, the real root of $2\lambda^2 + \lambda^{-4} = 3N$.

Then we assume that $W^*$, the electro-elastic energy of the membrane, is that of an ideal dielectric,

$$W^* = W^{ABB} - \frac{\varepsilon}{2}\boldsymbol{e}.\boldsymbol{e}, \tag{7}$$

where $\varepsilon$ is the electric permittivity of the material and $\boldsymbol{e}$ is the electric field, with magnitude $e = \frac{V}{h} = \frac{V}{H}\lambda^2$ (because $h = \lambda_r H = \lambda^2 H$).

The total potential energy of the DE balloon is thus $U_S = 4\pi R^2 H W^* + U_P$, or

$$U_S = 4\pi R^2 H\left\{\mu N\left[\frac{1}{6N}(2\lambda^2 + \lambda^{-4} - 3) - \ln(2\lambda^2 + \lambda^{-4} - 3N)\right] - \frac{\varepsilon}{2}\left(\frac{V}{H}\right)^2\lambda^4 - \frac{PR}{3H}\lambda^3\right\} + const. \tag{8}$$

The equation of motion is finally derived from the Euler-Lagrange equation,

$$\frac{d}{dt}\left(\frac{\partial \mathcal{L}}{\partial \dot\lambda}\right) - \left(\frac{\partial \mathcal{L}}{\partial \lambda}\right) = 0, \tag{9}$$

where $\mathcal{L} = U_K - U_S$ is the Lagrangian, as

$$\frac{\rho R^2}{\mu}\frac{d^2\lambda}{dt^2} + \frac{2}{3}(\lambda - \lambda^{-5}) - \frac{4N(\lambda - \lambda^{-5})}{2\lambda^2 + \lambda^{-4} - 3N} - \frac{2\varepsilon V^2}{\mu H^2}\lambda^3 - \frac{PR}{\mu H}\lambda^2 = 0. \tag{10}$$

We introduce the dimensionless measures of time $\tau = \frac{t}{R\sqrt{\rho/\mu_0}}$, voltage $\overline{V} = \frac{V}{H\sqrt{\mu_0/\varepsilon}}$, and inflation pressure $\overline{P} = \frac{PR}{\mu_0 H}$, and rewrite Eq. (10) as

$$\ddot\lambda + 2\left(\frac{N-1}{3N-1}\right)\left(\frac{2\lambda^2 + \lambda^{-4} - 9N}{2\lambda^2 + \lambda^{-4} - 3N}\right)(\lambda - \lambda^{-5}) - 2\overline{V}^2\lambda^3 - \overline{P}\lambda^2 = 0, \tag{11}$$

where the dot denotes differentiation with respect to $\tau$.

When the voltage is applied in a *Heaviside step*, from $\overline{V} = 0$ to the static voltage $\overline{V} = \overline{V}_{DC}$ at the initial time $\tau = 0$, the equation of motion becomes

$$\ddot\lambda + 2\left(\frac{N-1}{3N-1}\right)\left(\frac{2\lambda^2 + \lambda^{-4} - 9N}{2\lambda^2 + \lambda^{-4} - 3N}\right)(\lambda - \lambda^{-5}) - 2\overline{V}_{DC}^2\lambda^3 - \overline{P}\lambda^2 = 0. \tag{12}$$

When the voltage is *alternating*, say it is varying sinusoidally as

$$\overline{V} = \overline{V}_{DC} + \overline{V}_{AC}\sin(\Omega\tau), \tag{13}$$

where $\Omega$ denotes the nondimensional excitation frequency, then the non-dimensional AC dynamic equation is obtained as

$$\ddot{\lambda} + 2\left(\frac{N-1}{3N-1}\right)\left(\frac{2\lambda^2 + \lambda^{-4} - 9N}{2\lambda^2 + \lambda^{-4} - 3N}\right)(\lambda - \lambda^{-5}) - 2\overline{V}_{DC}^2\left[1 + \frac{\overline{V}_{AC}}{\overline{V}_{DC}}\sin(\Omega\tau)\right]^2 \lambda^3 - \overline{P}\lambda^2 = 0. \tag{14}$$

Finally, when put under an increasing *quasi-static* electrical load, the DE spherical membrane may undergo the static pull-in instability (when $N$ is large and the behaviour is close to that of a neo-Hookean material) and even the snap-through instability (when $N$ is not large and the strain-stiffening effect is marked). The voltage-stretch relationship for this loading is found by taking the time derivative in Eq. (14) to be identically zero:

$$2\left(\frac{N-1}{3N-1}\right)\left(\frac{2\lambda^2 + \lambda^{-4} - 9N}{2\lambda^2 + \lambda^{-4} - 3N}\right)(\lambda - \lambda^{-5}) - 2\overline{V}^2 \lambda^3 - \overline{P}\lambda^2 = 0, \tag{15}$$

from which the voltage follows as

$$\overline{V} = \sqrt{\left(\frac{N-1}{3N-1}\right)\left(\frac{2\lambda^2 + \lambda^{-4} - 9N}{2\lambda^2 + \lambda^{-4} - 3N}\right)(\lambda^{-2} - \lambda^{-8}) - \frac{1}{2}\overline{P}\lambda^{-1}}. \tag{16}$$

To find $\overline{V}_c$, the critical voltage of quasi-static instability, which occurs at the critical amount of stretch $\lambda_c$, we write that $\frac{d^2 U_S}{d\lambda^2} = 0$ [28], or

$$\frac{N-1}{3N-1}\left[\frac{24N(\lambda - \lambda^{-5})^2}{(2\lambda^2 + \lambda^{-4} - 3N)^2} + \frac{2\lambda^2 + \lambda^{-4} - 9N}{2\lambda^2 + \lambda^{-4} - 3N}(1 + 5\lambda^{-6})\right] - 3\overline{V}^2 \lambda^2 - \overline{P}\lambda = 0. \tag{17}$$

We then solve the two equations (15) and (17) simultaneously to obtain $\overline{V}_c$ and $\lambda_c$ at a given level of internal pressure $\overline{P}$. To find the location of all critical points, we may eliminate $\overline{P}$, so that they are on the curve

$$\overline{V} = \sqrt{\frac{N-1}{3N-1}\left[\frac{24N(1 - \lambda^{-6})^2}{(2\lambda^2 + \lambda^{-4} - 3N)^2} - \frac{2\lambda^2 + \lambda^{-4} - 9N}{2\lambda^2 + \lambda^{-4} - 3N}(\lambda^{-2} - 7\lambda^{-8})\right]}. \tag{18}$$

# 3 Stability analysis

## 3.1 Static pull-in and snap-through instabilities

By plotting the quasi-static voltage versus different values of the stretch according to Eq. (16), we identify the onset of static instabilities.

The resulting non-dimensional voltage-stretch $\overline{V} - \lambda$ curve is depicted in Fig. 2(a) for the $N = 30$ case. This value is representative of many polymers: for example, tensile experiments show that rubber vulcanizates have a maximum extension stretch $\lambda_m$ of about 10x, see [40]. In simple tension, $I_1 = \lambda^2 + 1/\lambda$, so that the constraint $I_1 < 3N$ when $N = 30$ gives $\lambda_m \sim 10$. Other materials used as dielectric elastomers, such as VHB 4905, an acrylic elastomer produced by 3M, seem to exhibit earlier strain-stiffening effect [55, 56]. See also Anssari-Benham et al. [52], who model the inflation of elastic rubber balloons with values of $N$ typically around 5, 15 and 30. For our study here, we take $N = 4, 15, 30$ in turn, corresponding to maximum stretch $\lambda_m = 2.45, 4.74, 6.71$, respectively.

The internal pressure is taken as $\overline{P} = 0.0, 0.4, 0.8, 1.2$. As depicted in Fig. 2(a), the voltage of static equilibrium increases sharply at first, and the radius increases moderately on a stable path (in the sense that $\frac{d^2 U_s}{d\lambda^2} > 0$). Then a maximum is reached, at the critical stretch $\lambda_c^S$ and critical voltage $\overline{V}_c^S$, indicated by a cross. At that stage the *snap-through instability* takes place: the membrane is expected to stretch dramatically at fixed voltage (as indicated by horizontal arrows), although this scenario is unlikely to unfold fully, as wrinkling instability or electrical breakdown will occur along the way [31]. The dotted line corresponds to the location of the instability points, obtained by Eq. (18).

In Fig. 2(b), we show the influence of the limiting-chain effect on the snap-through static instability, by plotting the same curves when $N = 4, 15, 30$, in turn. We see that balloons that are made of early-stiffening materials are more likely to be stable in static loading. For example, with some inflation pressure $\overline{P} > 0$, a balloon with $N = 4$ can avoid the snap-through static instability completely and be inflated smoothly by increasing voltage, up to stretch $\lambda \sim 2.4$.

The general trends found here for the quasi-static voltage versus stretch plots are consistent with those of previous studies based on other strain-stiffening models, such as that by Lv et al. [43] (using the Gent model) or that by Rudykh et al. [57] (using the Ogden model).

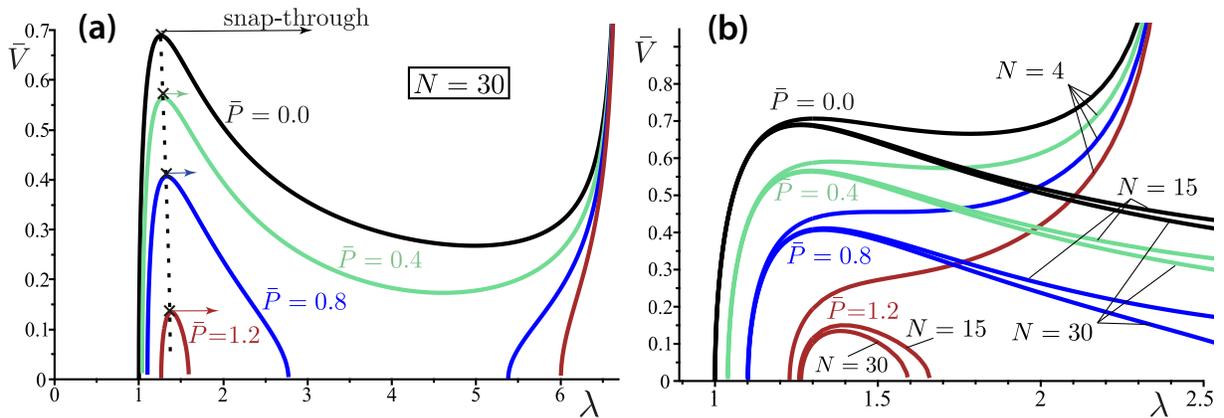

**Figure 2.** (a) Static instability in the DE balloon with $N = 30$, when it is under internal pressure $\overline{P} = 0.0, 0.4, 0.8, 1.2$. The crosses correspond to the critical points of static snap-through instability, at $(\lambda_c^S, \overline{V}_c^S)$ = (1.263, 0.6889), (1.293, 0.5639), (1.332, 0.4069), (1.385, 0.1347), respectively. (b) Influence of the strain-stiffening effect: as the number of Kuhn segments decreases: $N = 30, 15, 4$, the material stiffens earlier, which may lead to the disappearance of the snap-through instability, see $N = 4$ curves when $P > 0$.

## 3.2 Dynamic pull-in and snap-through instabilities

To study the *dynamic stability* of the balloon when it is inflated quasi-statically by inflation pressure $\bar{P}$ and then by a step voltage $H(t)\bar{V}_{DC}$, where $H$ is the Heaviside function, we follow the energy-based technique developed by Sharma et al. [28] (see also Khurana et al. [54]).

The Hamiltonian of the system is $\mathcal{H}(t) = U_K(t) + U_S(t)$, or, in its non-dimensional form,

$$\bar{\mathcal{H}}(\tau) = \tfrac{1}{2}\dot{\lambda}^2 + \tfrac{N-1}{3N-1}\left[\tfrac{1}{2}(2\lambda^2 + \lambda^{-4} - 3) - 3N \ln\left(\tfrac{2\lambda^2 + \lambda^{-4} - 3N}{3 - 3N}\right)\right] - \tfrac{1}{2}\bar{V}_{DC}^{\;2}\lambda^4 - \tfrac{1}{3}\bar{P}(\lambda^3 - 1). \tag{19}$$

At $\tau = 0$, the balloon is at rest: $\dot{\lambda}(0) = 0$ (and $U_K(0) = 0$), inflated by the pressure $\bar{P}$ to a stretch $\lambda(0) = \lambda_p$, found by solving Eq. (15) at $\bar{V} = 0$, that is, by solving

$$2\left(\tfrac{N-1}{3N-1}\right)\left(\tfrac{2\lambda_p^2 + \lambda_p^{-4} - 9N}{2\lambda_p^2 + \lambda_p^{-4} - 3N}\right)\left(\lambda_p - \lambda_p^{-5}\right) - \bar{P}\lambda_p^2 = 0. \tag{20}$$

Provided the balloon undergoes a periodic motion, at the time $\tilde{\tau}$ of maximum overshoot, we have $\dot{\lambda}(\tilde{\tau}) = 0$ (and $U_K(\tilde{\tau}) = 0$), and the stretch is $\tilde{\lambda}$, say.

The whole system is conservative, so that $\mathcal{D}(\tilde{\lambda}) = \bar{\mathcal{H}}(\tilde{\tau}) - \bar{\mathcal{H}}(0) = 0$, or

$$\mathcal{D}(\tilde{\lambda}) \equiv \tfrac{N-1}{3N-1}\left[\tfrac{1}{2}\left(2\tilde{\lambda}^2 - 2\lambda_p^2 + \tilde{\lambda}^{-4} - \lambda_p^{-4}\right) - 3N \ln\left(\tfrac{2\tilde{\lambda}^2 + \tilde{\lambda}^{-4} - 3N}{2\lambda_p^2 + \lambda_p^{-4} - 3N}\right)\right]$$
$$- \tfrac{1}{2}\bar{V}_{DC}^{\;2}(\tilde{\lambda}^4 - \lambda_p^4) - \tfrac{1}{3}\bar{P}(\tilde{\lambda}^3 - \lambda_p^3) = 0. \tag{21}$$

Hence, we obtain the dynamic loading curves by solving Eq. (20) and (21) simultaneously and plotting the corresponding $\bar{V}_{DC} - \tilde{\lambda}$ curve for a given inflation pressure $\bar{P}$.

To find the critical loads and stretches of dynamic instability, we write $\mathcal{D}'(\tilde{\lambda}) = 0$, which is Eq. (15) written at $\lambda = \tilde{\lambda}$. By solving this equation, simultaneously with Eq. (20) and (21), we obtain the critical parameters, as displayed in Table 1 for different levels of inflation pressure and number of Kuhn segments $N$. We see the clear trend that dynamic loadings allow for greater stretches to be attained, at lower voltages, than static loadings, in line with the results of Sharma et al. [28] for the neo-Hookean and Ogden models, and of Khurana et al. [54] for the vibrations of a plate. Moreover, we see that same trend when the material is made to stiffen earlier and earlier by decreasing $N$, with significant improvements in the values of the actuation stretch $\lambda_a = \lambda_c/\lambda_p$, and even suppression of the snap-through instability (in those latter cases, the actuation stretch is computed as $\lambda_a = \lambda_m/\lambda_p$.)

*Table 1: Critical stretches and voltages for the static and dynamic snap-through instabilities, for different inflation pressure levels and number of chain elements. The loading static stretches $\lambda_p$, maximum stretch $\lambda_m$, and actuation stretches $\lambda_a = \lambda_c/\lambda_p$ (or $\lambda_m/\lambda_p$ if there is no instability) are also given.*

|  | $N = 30$ ($\lambda_m = 6.078$) | | $N = 15$ ($\lambda_m = 4.743$) | | $N = 4$ ($\lambda_m = 2.447$) | |
|---|---|---|---|---|---|---|
| $\overline{P} = 0.0$ | $\lambda_p = 1.0$ | | $\lambda_p = 1.0$ | | $\lambda_p = 1.0$ | |
|  | $\overline{V_c^S} = 0.6889$ | $\overline{V_c^D} = 0.6489$ | $\overline{V_c^S} = 0.6907$ | $\overline{V_c^D} = 0.6511$ | $\overline{V_c^S} = 0.7066$ | no dynamic |
|  | $\lambda_c^S = 1.263$ | $\lambda_c^D = 1.472$ | $\lambda_c^S = 1.266$ | $\lambda_c^D = 1.481$ | $\lambda_c^S = 1.307$ | instability |
|  | $\lambda_a^S = 1.263$ | $\lambda_a^D = 1.472$ | $\lambda_a^S = 1.266$ | $\lambda_a^D = 1.481$ | $\lambda_a^S = 1.309$ | $\lambda_a^D = 2.447$ |
| $\overline{P} = 0.4$ | $\lambda_p = 1.040$ | | $\lambda_p = 1.040$ | | $\lambda_p = 1.039$ | |
|  | $\overline{V_c^S} = 0.5639$ | $\overline{V_c^D} = 0.5293$ | $\overline{V_c^S} = 0.5665$ | $\overline{V_c^D} = 0.5323$ | $\overline{V_c^S} = 0.5909$ | no dynamic |
|  | $\lambda_c^S = 1.293$ | $\lambda_c^D = 1.491$ | $\lambda_c^S = 1.298$ | $\lambda_c^D = 1.501$ | $\lambda_c^S = 1.361$ | instability |
|  | $\lambda_a^S = 1.244$ | $\lambda_a^D = 1.434$ | $\lambda_a^S = 1.249$ | $\lambda_a^D = 1.444$ | $\lambda_a^S = 1.309$ | $\lambda_a^D = 2.355$ |
| $\overline{P} = 0.8$ | $\lambda_p = 1.101$ | | $\lambda_p = 1.101$ | | $\lambda_p = 1.100$ | |
|  | $\overline{V_c^S} = 0.4069$ | $\overline{V_c^D} = 0.3790$ | $\overline{V_c^S} = 0.4112$ | $\overline{V_c^D} = 0.3837$ | $\overline{V_c^S} = 0.4553$ | no dynamic |
|  | $\lambda_c^S = 1.332$ | $\lambda_c^D = 1.503$ | $\lambda_c^S = 1.339$ | $\lambda_c^D = 1.517$ | $\lambda_c^S = 1.541$ | instability |
|  | $\lambda_a^S = 1.210$ | $\lambda_a^D = 1.365$ | $\lambda_a^S = 1.216$ | $\lambda_a^D = 1.378$ | $\lambda_a^S = 1.401$ | $\lambda_a^D = 2.224$ |
| $\overline{P} = 1.2$ | $\lambda_p = 1.264$ | | $\lambda_p = 1.258$ | | $\lambda_p = 1.229$ | |
|  | $\overline{V_c^S} = 0.1347$ | $\overline{V_c^D} = 0.1214$ | $\overline{V_c^S} = 0.1504$ | $\overline{V_c^D} = 0.1362$ | no static | no dynamic |
|  | $\lambda_c^S = 1.385$ | $\lambda_c^D = 1.459$ | $\lambda_c^S = 1.396$ | $\lambda_c^D = 1.483$ | instability | instability |
|  | $\lambda_a^S = 1.096$ | $\lambda_a^D = 1.154$ | $\lambda_a^S = 1.110$ | $\lambda_a^D = 1.179$ | $\lambda_a^S = 1.991$ | $\lambda_a^D = 1.991$ |

We may then compare the static and dynamic stability curves and study how they are influenced by the number of Kuhn segments $N$. Figure 3 presents representative curves of what happens when the material exhibits early strain-stiffening effect: it compares the $N = 4$ case to the $N = 15$ case, when the balloon is pre-inflated by pressure $\overline{P} = 0.8$. It clearly shows the delay (or vanishing) in the onset of snap-through instability when going from static to dynamic loading, as described above.

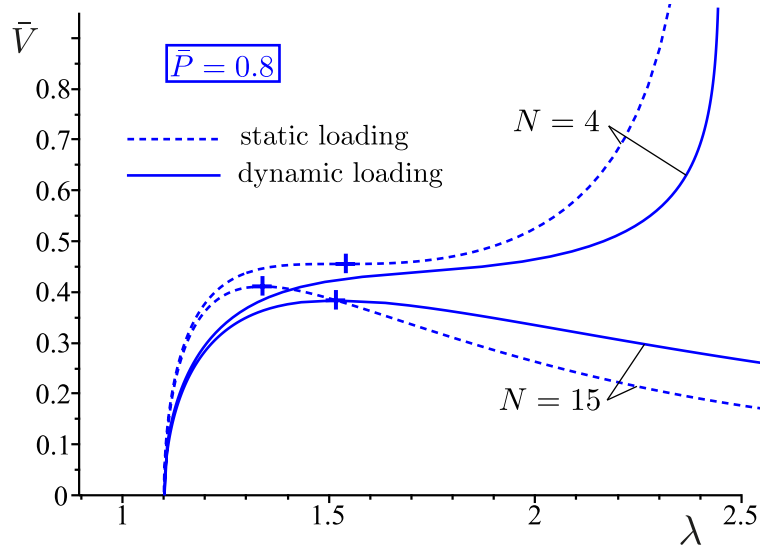

**Figure 3. Comparing static to dynamic loading for a DE balloon with early ($N = 4$) or late ($N = 15$) strain-stiffening effect, when it is under internal pressure $\overline{P} = 0.8$. The crosses correspond to the critical points of snap-through instability (values in Table 1). Clearly, dynamic loading allows for greater stretching before losing stability, requiring a lower voltage, and even for complete avoidance of snap-through instability for the free vibrations generated by a step continuous voltage.**

## 4 Nonlinear vibrations and motions

In this section we analyse the dynamic response of the balloon. First, we present results when the system is excited by a DC Heaviside step voltage signal, and then when it is under an AC time-dependent voltage. To solve the governing equations, we use the Runge-Kutta method, neglecting possible transient responses by a long-time integration. Using time-integration, we generate time-stretch histories, phase-plane diagrams, Poincaré sections, and fast Fourier transforms (FFT) plots, and identify different dynamic motion scenarios: periodicity, quasi-periodicity, and chaos.

### 4.1 DC dynamic instability

Here we investigate the influence of the number of Kuhn segments $N$ on the DC dynamic instability. In other words, we illustrate and confirm the results summarised in Table 1 by an example.

In Fig. 4 we compare the possible onset of DC dynamic instability for two values of $N$ ($N = 4, 15$) when the balloon is under inflation pressure $\bar{P} = 0.8$.

When the voltage is below dynamic instability thresholds, say $\bar{V}_{DC} = 0.3836$ (see Table 1), we clearly see that the balloon undergoes periodic motions: in the time-stretch diagrams (Fig. 4a), regular and predictable trajectories are observed.

Just above the threshold $\overline{V_c^D} = 0.3837$ of dynamic instability for the softer membrane with $N = 15$, at $\bar{V}_{DC} = 0.3838$, say, the balloon can no longer sustain periodic motions and the stretch increases uncontrollably, reflecting dynamic instability and confirming the predictions of the previous section.

In contrast, we can increase the voltage for the stiffer membrane with $N = 4$ to high values, $\bar{V}_{DC} = 0.7$, say, and still observe periodic motions, even though the voltage is beyond the threshold of quasi-static instability (which is $\overline{V_c^S} = 0.4553$, see Table 1). These various behaviours are also reflected in the corresponding phase diagrams of Fig. 4b.

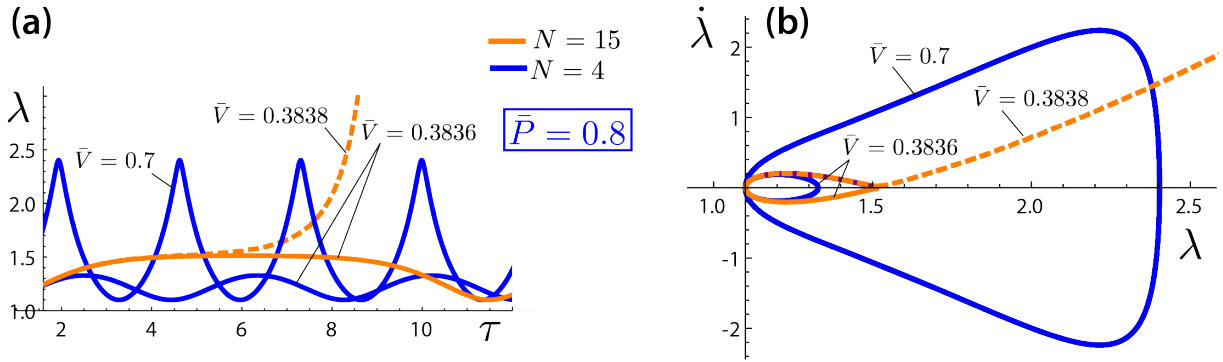

**Figure 4.** The DC dynamic response of the DE balloon for $\bar{P} = 0.8$, when the number of Kuhn segments is $N = 4$ (early strain-stiffening effect) and $N = 15$ (late strain-stiffening effect). In the latter case, as soon as the voltage exceeds the critical voltage threshold $\bar{V}_{DC} = 0.3837$, the motion loses its periodicity. In the former case, there is no critical voltage of dynamic instability, even though there is one for static instability.

The amplitude of the free nonlinear oscillations is $\frac{1}{2}(\tilde{\lambda} - \lambda_p)$, where we recall that $\tilde{\lambda}$, the maximum overshoot, is computed from Eq. (21). Then, writing that $\bar{\mathcal{H}}(\tau) = \bar{\mathcal{H}}(0)$ (which is true for all $\tau$ because the system is conservative), gives $\dot{\lambda}^2 = -2D(\lambda)$, see Eqs. (19) and (21). It follows that the period $T$ of the oscillations is given by the integral $T = \sqrt{2}\int_{\lambda_p}^{\tilde{\lambda}} \frac{d\lambda}{\sqrt{-D(\lambda)}}$. Hence, when $\overline{V}_{DC} = 0.3836, 0.7$, we find $\tilde{\lambda} = 1.339, 2.403$, and $T = 3.904, 2.685$, respectively, for the early stiffening elastomer with $N = 4$, see Fig. 4.

**4.2 AC dynamic instability**

One of the important phenomena that has been noted in DEs is the possible emergence of chaos from forced oscillations. In general, chaos may arise in a system due to sensitivity to the initial condition and/or a change in one or several of the system's parameters. In this section, the influence of parameter $N$ on the nonlinear vibrations of the DE balloon is studied with special consideration to the possibility of chaos.

For illustrative and representative purposes, we take the applied inflation pressure to be $\overline{P} = 0.8$, initial stretch at rest taken from Table 1, i.e., $\lambda(0) = \lambda_p = 1.1, \dot{\lambda}(0) = 0$, and we chose the excitation frequency as $\Omega = 1.5$.

Fig. 5 illustrates the nonlinear vibrations of an electroactive balloon with early-stiffening characteristics $N = 4$, put under dynamic step voltage loading $\overline{V}_{DC} = 0.3836$, and voltage forcing with amplitude $\overline{V}_{AC}/\overline{V}_{DC} = 0.1$. We clearly see the emergence of beats after a certain time (once the transient response has died out) in the time-stretch plot. The quasi-periodicity attractor is observed in this figure by inter-operating the phase portrait, the Poincaré section, and FFT. The Poincaré section is drawn by sectioning the phase portrait at every period of the forcing frequency $\Omega$ (we left out enough periods to get rid of the transient response). The Poincaré section indicates a closed curve, the sign of quasi-periodic oscillations. Separated spectra in FFT also confirm quasi-periodicity.

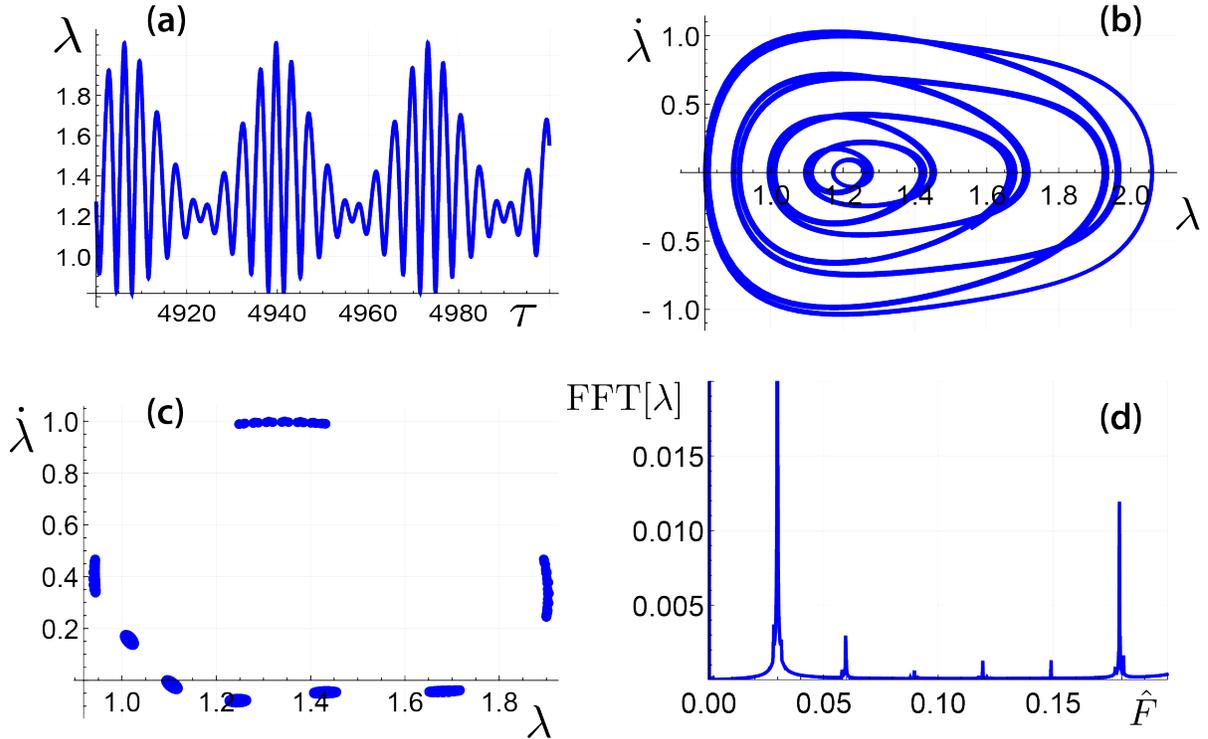

**Figure 5. Quasiperiodic vibrations of a DE balloon made of early-stiffening material with $N = 4$, for $\overline{P} = 0.8$, $\overline{V}_{DC} = 0.3836, \overline{V}_{AC}/\overline{V}_{DC} = 0.1$. (a) Time-stretch diagram, (b) Phase portrait, (c) Poincaré section, (d) Fast Fourier Transform.**

Changing now from $N = 4$ to a softer material with $N = 15$, while keeping the other parameters the same, we observe a change from quasi-periodicity to chaos attractor, see Fig. 6. This change was to be expected, because the voltage loading $\overline{V}_{DC} = 0.3836$ is very close to the critical value of dynamic instability ($\overline{V}_c = 0.3837$), and the addition of the alternative voltage increases the magnitude of the voltage by 10%, enough to trigger instability and drift toward large actuation, which will then hit an upper ceiling at $\lambda_m = 4.743$.

Hence in Figs. 6ab, the time trace and phase plane reveal irregular and unpredictable trajectories, indicating chaotic behaviour, and all bonded above at $\lambda_m = 4.743$. An infinite number of points can be generated in the Poincaré section (Fig. 6c), showing the complex dynamical behaviour of the balloon. The FFT diagram illustrates a continuous distribution and infinite number of spectra, confirming the chaos phenomenon.

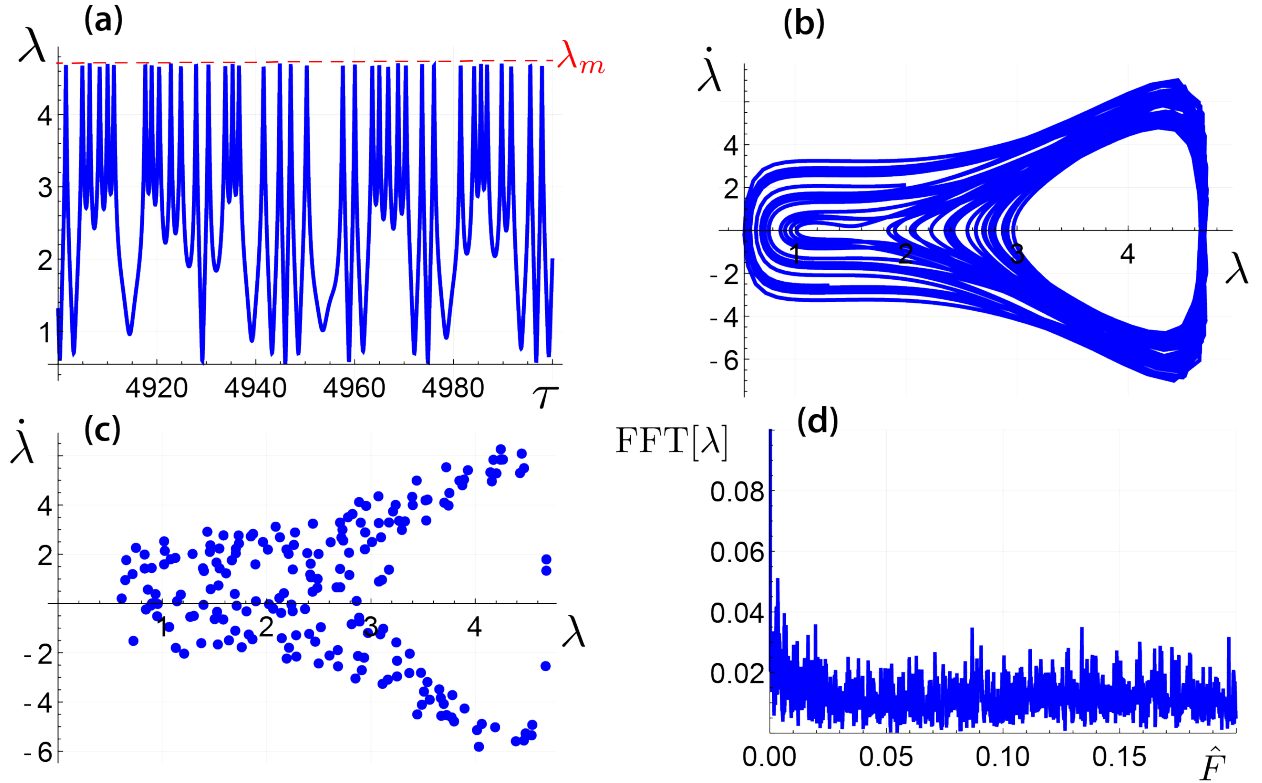

**Figure 6. Chaotic vibrations of the softer DE balloon with $N = 15$, and inflation pressure $\overline{P} = 0.8$, when the DC voltage $\overline{V}_{DC} = 0.3836$, is too close to the critical voltage of dynamic instability. Here the superposed AC voltage $\overline{V}_{AC} = 0.1\overline{V}_{DC}$ has enough amplitude to send the voltage above the threshold of instability. (a) Time-stretch diagram and (b) Phase portrait, showing erratic vibrations, all bonded above by the limiting stretch $\lambda_m = 4.743$. (c) Poincaré section and (d) FFT, confirming the chaotic behaviour.**

When we decrease the DC voltage in Fig. 6 to $\overline{V}_{DC} = 0.3$ for the $N = 15$ model, we place the motion sufficiently away from the critical voltage of dynamic stability to observe that the response of the system becomes quasi-periodic. The resulting numerical integration and plots are very similar to those of Fig. 5 and are not reproduced here. This means that for a weak strain-stiffening DE balloon, we can reach a stable quasiperiodic vibration by decreasing the DC voltage. However, the static voltage must be much smaller than the critical static and DC dynamic voltages (see Table. 1).

We then look at what happens to forced AC oscillations for the DE balloon with strong strain-stiffening effect ($N = 4$), when the DC static voltage is large enough to bring the stretch close to the limiting stretch $\lambda_m = 2.447$. Hence we take $\overline{V}_{DC} = 0.7$, which according to the results of Fig. 3 generates a maximum overshoot $\tilde{\lambda} = 2.403$. This value is within less than of 2% of $\lambda_m$, and superposing an AC voltage will generate nonlinear vibrations which hit that barrier and eventually lead to chaos, see Fig. 7.

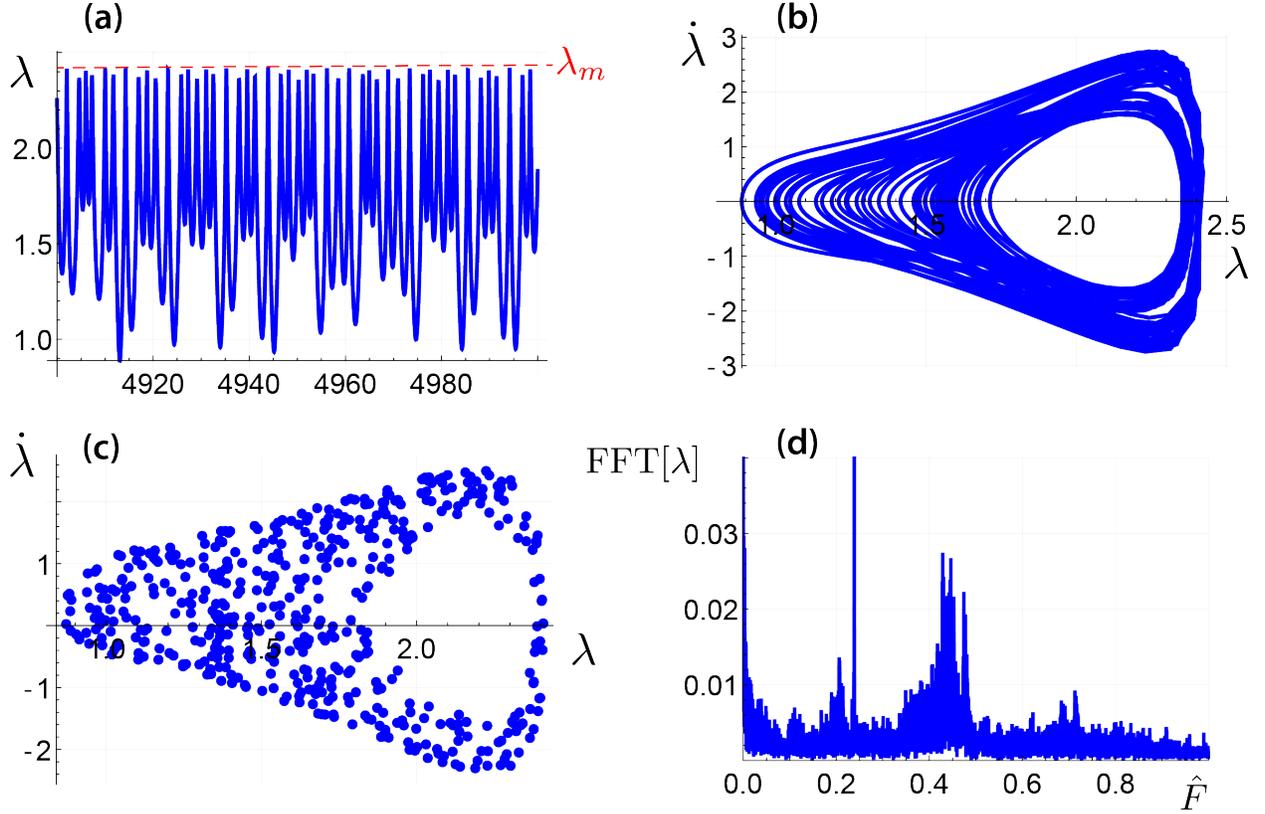

**Figure 7.** Chaotic vibrations of the DE balloon for $\overline{P} = 0.8, \overline{V}_{DC} = 0.7, \frac{\overline{V}_{AC}}{\overline{V}_{DC}} = 0.1$ and $N = 4$. (a) Time-stretch diagram, (b) Phase portrait, (c) Poincaré section, (d) FFT.

In this section we focused on the role that $N$ might play in the occurrence or suppression of chaos, compared to very soft models such as the neo-Hookean model. We saw that a great variety of situations can arise, and presented some examples. For a more complete parametric study, see the work of Chen et al.[42].

## 5 Conclusion

We investigated nonlinear vibrations and instability modes of an inflated dielectric elastomer balloon modelled by a generalised neo-Hookean model with a strain-stiffening effect that captures the microscopic properties of elastomers. We derived the governing equations for two types of voltage loading: a static voltage (either applied quasi-statically or as a step function) and a sinusoidal voltage superposed on top of a step DC voltage.

We found and solved the equations for the static and dynamic snap-through instability modes. We then solved the dynamic vibration equations numerically by the Runge-Kutta method. The analysis was conducted by plotting and commenting the time-stretch responses, phase-plane diagrams, Poincaré sections, Fast Fourier Transforms, and stretch-voltage diagrams.

The main outcomes of the paper are as follows. (1) The microscopic properties of the DE balloon material affect greatly the instability and vibration modes. (2) As the number of links in a single chain *N* is decreased, the DE balloon material exhibits earlier strain-stiffening effects and presents greater static and dynamic stability. (3) Moreover, a decreasing Kuhn number *N* can suppress the chaotic oscillations found in softer materials (close to the neo-Hookean material), provided the stretch is not too close to the limiting-chain stretch.

The results presented in this paper cover the hyperelastic region of DEs. However, experimental tests have shown that DEs can also undergo inelastic (plastic) deformations, on account of different external factors such as the stress-controlled or voltage-controlled cyclic loading conditions, which is considered as a kind of ratchetting [58]. The analysis of this state of deformation is beyond the scope of the paper.

**Acknowledgements**
This work is supported by the Seagull Program of Zhejiang Province, the National Natural Science Foundation of China (No. 11872329), the Natural Science Foundation of Zhejiang Province (No. LD21A020001), and the 111 Project (No. B21034). The support from the Shenzhen Scientific and Technological Fund for R&D, PR China (No. 2021Szvup152) is also acknowledged.

**References**

1. Sato, J., Sekine, T., Yi-Fei, W., Takeda, Y., Matsui, H., Kumaki, D., Santos, F.D. Dos, Miyabo, A., Tokito, S.: Ferroelectric polymer-based fully printed flexible strain rate sensors and their application for human motion capture. Sensors Actuators, A Phys. 295, (2019). https://doi.org/10.1016/j.sna.2019.05.022
2. Jiang, Y., Zhou, M., Shen, Z., Zhang, X., Pan, H., Lin, Y.H.: Ferroelectric polymers and their nanocomposites for dielectric energy storage applications, (2021)
3. Liu, Y., Haibibu, A., Xu, W., Han, Z., Wang, Q.: Observation of a Negative Thermal Hysteresis in Relaxor Ferroelectric Polymers. Adv. Funct. Mater. 30, (2020). https://doi.org/10.1002/adfm.202000648
4. Li, Q., Wang, Q.: Ferroelectric Polymers and Their Energy-Related Applications. Macromol. Chem. Phys. 217, (2016). https://doi.org/10.1002/macp.201500503
5. Liu, Y., Yang, T., Zhang, B., Williams, T., Lin, Y.T., Li, L., Zhou, Y., Lu, W., Kim, S.H., Chen, L.Q., Bernholc, J., Wang, Q.: Structural Insight in the Interfacial Effect in Ferroelectric Polymer Nanocomposites. Adv. Mater. 32, (2020). https://doi.org/10.1002/adma.202005431
6. Chen, X., Han, X., Shen, Q.D.: PVDF-Based Ferroelectric Polymers in Modern Flexible Electronics, (2017)
7. Li, H., Wang, R., Han, S.T., Zhou, Y.: Ferroelectric polymers for non-volatile memory devices: a review, (2020)
8. Su, J., Harrison, J.S., St. Clair, T.L., Bar-Cohen, Y., Leary, S.: Electrostrictive graft elastomers and applications. In: Materials Research Society Symposium - Proceedings (2000)
9. Wang, Y., Sun, C., Zhou, E., Su, J.: Deformation mechanisms of electrostrictive graft elastomer. Smart Mater. Struct. 13, (2004). https://doi.org/10.1088/0964-1726/13/6/011
10. Davidson, J.R., Krebs, H.I.: An Electrorheological Fluid Actuator for Rehabilitation Robotics. IEEE/ASME Trans. Mechatronics. 23, (2018). https://doi.org/10.1109/TMECH.2018.2869126
11. Xu, Z., Wu, H., Zhang, M., Wu, J., Wen, W.: The research progress of electrorheological fluids, (2017)
12. Sheng, P., Wen, W.: Electrorheological fluids: Mechanisms, dynamics, and microfluidics applications. Annu. Rev. Fluid Mech. 44, (2011). https://doi.org/10.1146/annurev-fluid-120710-101024
13. Kuznetsov, N.M., Belousov, S.I., Kamyshinsky, R.A., Vasiliev, A.L., Chvalun, S.N., Yudina, E.B., Ya. Vul, A.: Detonation nanodiamonds dispersed in polydimethylsiloxane as a novel electrorheological fluid: Effect of nanodiamonds surface. Carbon N. Y. 174, (2021). https://doi.org/10.1016/j.carbon.2020.12.014
14. Lu, Q., Han, W.J., Choi, H.J.: Smart and functional conducting polymers: Application to electrorheological fluids. Molecules. 23, (2018). https://doi.org/10.3390/molecules23112854
15. Biswal, D.K., Bandopadhya, D., Dwivedy, S.K.: A non-linear dynamic model of ionic polymer-metal composite (IPMC) cantilever actuator. Int. J. Automot. Mech. Eng. 16, (2019). https://doi.org/10.15282/ijame.16.1.2019.17.0479
16. Leronni, A., Bardella, L.: Modeling actuation and sensing in ionic polymer metal composites by electrochemo-poromechanics. J. Mech. Phys. Solids. 148, (2021). https://doi.org/10.1016/j.jmps.2021.104292
17. Kweon, B.C., Sohn, J.S., Ryu, Y., Cha, S.W.: Energy harvesting of ionic polymer-metal composites based on microcellular foamed nafion in aqueous environment. Actuators. 9, (2020). https://doi.org/10.3390/ACT9030071
18. Ma, S., Zhang, Y., Liang, Y., Ren, L., Tian, W., Ren, L.: High-Performance Ionic-Polymer–Metal Composite: Toward Large-Deformation Fast-Response Artificial Muscles. Adv. Funct. Mater. 30, (2020). https://doi.org/10.1002/adfm.201908508
19. Sideris, E.A., De Lange, H.C., Hunt, A.: An Ionic Polymer Metal Composite (IPMC)-Driven Linear Peristaltic Microfluidic Pump. IEEE Robot. Autom. Lett. 5, (2020). https://doi.org/10.1109/LRA.2020.3015452
20. Yang, L., Zhang, D., Wang, H., Zhang, X.: Actuation Modeling of Ionic–Polymer Metal Composite Actuators Using Micromechanics Approach. Adv. Eng. Mater. 22, (2020). https://doi.org/10.1002/adem.202000537
21. Ahn, S.K., Kasi, R.M., Kim, S.C., Sharma, N., Zhou, Y.: Stimuli-responsive polymer gels. Soft Matter. 4, (2008). https://doi.org/10.1039/b714376a
22. Goponenko, A. V., Dzenis, Y.A.: Role of mechanical factors in applications of stimuli-responsive polymer gels – Status and prospects, (2016)
23. Haraguchi, K.: Stimuli-responsive nanocomposite gels. Colloid Polym. Sci. 289, (2011). https://doi.org/10.1007/s00396-010-2373-9
24. Echeverria, C., Fernandes, S.N., Godinho, M.H., Borges, J.P., Soares, P.I.P.: Functional stimuli-responsive gels: Hydrogels and microgels. Gels. 4, (2018). https://doi.org/10.3390/gels4020054
25. Yang, X., Zhang, G., Zhang, D.: Stimuli responsive gels based on low molecular weight gelators. J. Mater. Chem. 22, (2012). https://doi.org/10.1039/c1jm13205a
26. Zurlo, G., Destrade, M., DeTommasi, D., Puglisi, G.: Catastrophic Thinning of Dielectric Elastomers. Phys. Rev. Lett. 118, (2017). https://doi.org/10.1103/PhysRevLett.118.078001
27. Zurlo, G., Destrade, M., Lu, T.: Fine tuning the electro-mechanical response of dielectric elastomers. Appl. Phys. Lett. 113, (2018). https://doi.org/10.1063/1.5053643
28. Sharma, A.K., Arora, N., Joglekar, M.M.: DC dynamic pull-in instability of a dielectric elastomer balloon: An energy-based approach. In: Proceedings of the Royal Society A: Mathematical, Physical and Engineering Sciences (2018)
29. Ghosh, A., Basu, S.: Soft dielectric elastomer tubes in an electric field. J. Mech. Phys. Solids. 150, (2021). https://doi.org/10.1016/j.jmps.2021.104371
30. Alibakhshi, A., Heidari, H.: Nonlinear dynamic responses of electrically actuated dielectric elastomer-based microbeam resonators. J. Intell. Mater. Syst. Struct. (2021). https://doi.org/10.1177/1045389x211023584
31. Su, Y., Conroy Broderick, H., Chen, W., Destrade, M.: Wrinkles in soft dielectric plates. J. Mech. Phys. Solids. 119, (2018). https://doi.org/10.1016/j.jmps.2018.07.001
32. Su, Y., Chen, W., Dorfmann, L., Destrade, M.: The effect of an exterior electric field on the instability of dielectric plates. Proc. R. Soc. A Math. Phys. Eng. Sci. 476, (2020). https://doi.org/10.1098/rspa.2020.0267
33. Zhu, J., Cai, S., Suo, Z.: Nonlinear oscillation of a dielectric elastomer balloon. Polym. Int. 59, 378–383 (2010). https://doi.org/10.1002/PI.2767

34. Yong, H., He, X., Zhou, Y.: Dynamics of a thick-walled dielectric elastomer spherical shell. Int. J. Eng. Sci. 49, (2011). https://doi.org/10.1016/j.ijengsci.2011.03.006
35. Jin, X., Huang, Z.: Random response of dielectric elastomer balloon to electrical or mechanical perturbation. J. Intell. Mater. Syst. Struct. 28, (2017). https://doi.org/10.1177/1045389X16649446
36. Alibakhshi, A., Heidari, H.: Analytical approximation solutions of a dielectric elastomer balloon using the multiple scales method. Eur. J. Mech. A/Solids. 74, (2019). https://doi.org/10.1016/j.euromechsol.2019.01.009
37. Tang, D., Lim, C.W., Hong, L., Jiang, J., Lai, S.K.: Analytical asymptotic approximations for large amplitude nonlinear free vibration of a dielectric elastomer balloon. Nonlinear Dyn. 88, (2017). https://doi.org/10.1007/s11071-017-3374-8
38. Liu, F., Zhou, J.: Shooting and Arc-Length Continuation Method for Periodic Solution and Bifurcation of Nonlinear Oscillation of Viscoelastic Dielectric Elastomers. J. Appl. Mech. Trans. ASME. 85, (2018). https://doi.org/10.1115/1.4038327
39. Tang, D., Lim, C.W., Hong, L., Jiang, J., Lai, S.K.: Dynamic Response and Stability Analysis with Newton Harmonic Balance Method for Nonlinear Oscillating Dielectric Elastomer Balloons. Int. J. Struct. Stab. Dyn. 18, (2018). https://doi.org/10.1142/S0219455418501523
40. Gent, A.N.: A new constitutive relation for rubber. Rubber Chem. Technol. 69, (1996). https://doi.org/10.5254/1.3538357
41. Chen, F., Wang, M.Y.: Dynamic performance of a dielectric elastomer balloon actuator. Meccanica. 50, (2015). https://doi.org/10.1007/s11012-015-0206-0
42. Chen, F., Zhu, J., Wang, M.Y.: Dynamic electromechanical instability of a dielectric elastomer balloon. EPL. 112, (2015). https://doi.org/10.1209/0295-5075/112/47003
43. Lv, X., Liu, L., Liu, Y., Leng, J.: Dynamic performance of dielectric elastomer balloon incorporating stiffening and damping effect. Smart Mater. Struct. 27, (2018). https://doi.org/10.1088/1361-665X/aab9db
44. Deng, Z.: Dynamic analysis and active control of a dielectric elastomer balloon covered by a protective passive layer. J. Mech. Eng. Res. 2, (2019). https://doi.org/10.30564/jmer.v2i1.914
45. Mao, R., Wu, B., Carrera, E., Chen, W.: Electrostatically tunable small-amplitude free vibrations of pressurized electro-active spherical balloons. Int. J. Non. Linear. Mech. 117, (2019). https://doi.org/10.1016/j.ijnonlinmec.2019.103237
46. Liang, X., Cai, S.: New electromechanical instability modes in dielectric elastomer balloons. Int. J. Solids Struct. 132–133, (2018). https://doi.org/10.1016/j.ijsolstr.2017.09.021
47. Arruda, E.M., Boyce, M.C.: A three-dimensional constitutive model for the large stretch behavior of rubber elastic materials. J. Mech. Phys. Solids. 41, (1993). https://doi.org/10.1016/0022-5096(93)90013-6
48. Itskov, M., Khiêm, V.N., Waluyo, S.: Electroelasticity of dielectric elastomers based on molecular chain statistics. Math. Mech. Solids. 24, (2019). https://doi.org/10.1177/1081286518755846
49. Pucci, E., Saccomandi, G.: A note on the gent model for rubber-like materials. Rubber Chem. Technol. 75, (2002). https://doi.org/10.5254/1.3547687
50. Mangan, R., Destrade, M.: Gent models for the inflation of spherical balloons. Int. J. Non. Linear. Mech. 68, (2015). https://doi.org/10.1016/j.ijnonlinmec.2014.05.016
51. Alibakhshi, A., Heidari, H.: Nonlinear dynamics of dielectric elastomer balloons based on the Gent-Gent hyperelastic model. Eur. J. Mech. A/Solids. 82, (2020). https://doi.org/10.1016/j.euromechsol.2020.103986
52. Anssari-Benam, A., Bucchi, A.: A generalised neo-Hookean strain energy function for application to the finite deformation of elastomers. Int. J. Non. Linear. Mech. 128, (2021). https://doi.org/10.1016/j.ijnonlinmec.2020.103626
53. Davidson, J.D., Goulbourne, N.C.: A nonaffine network model for elastomers undergoing finite deformations. J. Mech. Phys. Solids. 61, (2013). https://doi.org/10.1016/j.jmps.2013.03.009
54. Khurana, A., Kumar, D., Sharma, A.K., Joglekar, M.M.: Static and dynamic instability modeling of electro-magneto-active polymers with various entanglements and crosslinks. Int. J. Non. Linear.

Mech. 103865 (2021). https://doi.org/10.1016/J.IJNONLINMEC.2021.103865
55. Horgan, C.O.: A note on a class of generalized neo-Hookean models for isotropic incompressible hyperelastic materials. Int. J. Non. Linear. Mech. 129, (2021). https://doi.org/10.1016/j.ijnonlinmec.2020.103665
56. Anssari-Benam, A., Bucchi, A., Saccomandi, G.: Modelling the Inflation and Elastic Instabilities of Rubber-Like Spherical and Cylindrical Shells Using a New Generalised Neo-Hookean Strain Energy Function. J. Elast. (2021). https://doi.org/10.1007/s10659-021-09823-x
57. Rudykh, S., Bhattacharya, K., Debotton, G.: Snap-through actuation of thick-wall electroactive balloons. Int. J. Non. Linear. Mech. 47, (2012). https://doi.org/10.1016/j.ijnonlinmec.2011.05.006
58. Chen, Y., Kang, G., Yuan, J., Hu, Y., Li, T., Qu, S.: An electro-mechanically coupled visco-hyperelastic-plastic constitutive model for cyclic deformation of dielectric elastomers. Mech. Mater. 150, (2020). https://doi.org/10.1016/j.mechmat.2020.103575